\documentclass[11pt]{article}
\usepackage[dvips]{graphicx}
\usepackage{latexsym}
\usepackage{amssymb}
\newcommand{\MMM}{\mathbb{M}}
\newcommand{\C}{\mathbb{C}}

\newcommand{\R}{\mathbb{R}}

\begin{document}
\title{From 2D Integrable Systems \\
to Self-Dual Gravity}
\author{M. Dunajski \thanks{Work partly supported by \L{\'{o}}d\'{z}
University grant no. 457} 
, L. J. Mason and N. M. J. Woodhouse\\ The Mathematical Institute\\
24-29 St Giles, Oxford OX1 3LB, UK} \date{} 
\maketitle 

\abstract {We explain how to construct solutions to the self-dual
Einstein vacuum equations from solutions of various two-dimensional
integrable systems by exploiting the fact that the Lax formulations of
both systems can be embedded in that of the self-dual Yang--Mills
equations. We illustrate this by constructing explicit  self-dual vacuum
metrics on $\R^2\times \Sigma$, where $\Sigma$ is a homogeneous space for a
real subgroup of $SL(2, \C)$ associated with the two-dimensional system.}



\def\be{\begin{equation}}
\def\ee{\end{equation}}
\def\theequation{\thesection.\arabic{equation}}
\def\ss{sdiff($\Sigma^2$)}
\def\SS{SDiff($\Sigma^2$)}
\def\SM{SDiff ($M^4$)}
\def\ot{\otimes}
\def\xm{\mid \xi \mid}
\def\xib{\overline{\xi}}
\def\xit{\tilde{\xi}}
\def\Sm{\Sigma}
\def\Smt{\widetilde {\Sm}}
\def\s{\sigma} 
\def\Om{\Omega}
\def\os{\Om _{\Sm}}
\def\om{\omega}
\def\ov{\overline}
\def\pst{\widetilde {\phi}}
\def\ps{\phi}
\def\psm{\mid \ps \mid}
\def\psb{\ov{\ps}}
\def\ds{d_{\Sigma}}
\def\macierz{
\left (
\begin{array}{cccc}
0&1&0&0\\
1&0&0&0\\
0&0&0&1\\
0&0&1&0
\end{array}
\right )
}
\def\AZ{
\frac{1}{4}
\left (
\begin{array}{cc}
-{\phi}_x&-2{\sin}({\phi}/2)\\
2{\sin}({\phi}/2)&{\phi}_x
\end{array}
\right )
 }
\def\AW{
\frac{1}{4}
\left (
\begin{array}{cc}
-{\phi}_x&2{\sin}({\phi}/2)\\
-2{\sin}({\phi}/2)&{\phi}_x
\end{array}
\right )
}
\def\AU{
\frac{1}{4}
\left (
\begin{array}{cc}
{\phi}_t&-2{\cos}({\phi}/2)\\
-2{\cos}({\phi}/2)&-{\phi}_t
\end{array}
\right )
}
\def\AV{
\frac{1}{4}
\left (
\begin{array}{cc}
{\phi}_t&2{\cos}({\phi}/2)\\
2{\cos}({\phi}/2)&-{\phi}_t
\end{array}
\right )
}
\def\pau1{
\left (
\begin{array}{cc}
0&i\\
i&0
\end{array}
\right ) 
}
\def\lam{
\left (
\begin{array}{cc}
0&1\\
-1&0
\end{array}
\right ) 
}
\def\lap{
\left (
\begin{array}{cc}
i&0\\
0&-i
\end{array}
\right ) 
}
\def\Bk{
\left (
\begin{array}{cc}
0&0\\
-1&0
\end{array}
\right ) 
}
\def\B2{
\left (
\begin{array}{cc}
1&0\\
0&-1
\end{array}
\right ) 
}
\def\Ak{
\left (
\begin{array}{cc}
q&1\\
b&-q
\end{array}
\right ) 
}
\def\Qk{
\left (
\begin{array}{cc}
b_x&-2q_x\\
2w&-b_x
\end{array}
\right ) 
}
\def\A2{
\left (
\begin{array}{cc}
0&{\ps}\\
\mp {\psb}&0
\end{array}
\right ) 
}
\def\Q2{
\left (
\begin{array}{cc}
{\psm}^2&\ \pm \ps _x\\
\psb _x&-{\psm}^2
\end{array}
\right ) 
}
\def\M{
\left (
\begin{array}{cc}
A&B\\
C&D
\end{array}
\right ) 
}
\def\N{
\left (
\begin{array}{cc}
{\alpha}&{\beta}\\
{\ov{\beta}}&{\ov{\alpha}}
\end{array}
\right ) 
}
\def\i1{
\left (
\begin{array}{cc}
1&i\\
i&1
\end{array}
\right ) 
}
\def\mn{
\left (
\begin{array}{cc}
m\\
n
\end{array}
\right )
}
\def\mnt{
\left (
\begin{array}{cc}
m+t(am+bm)\\
n+t(cm-an)
\end{array}
\right )
}
\def\AM{
\left (
\begin{array}{cc}
a&b\\
c&-a
\end{array}
\right ) 
}
\def\Wt{\widetilde W}
\def\Zt{\widetilde Z}
\def\wt{\tilde w}
\def\zt{\tilde z}
\def\w{w}
\def\con{\,\rule{5.5pt}{0.3pt}\rule{0.3pt}{5.5pt}\,}
\def\p{\partial}  
\def\a{\alpha}
\def\at{\tilde \a}
\tolerance=400
\section{Introduction}

Ward \cite{Wa85} has observed that many integrable systems,
particularly in two dimensions, may be
obtained from the self-dual Yang--Mills (SDYM) equations by symmetry
reduction. See \cite{AC92} for a survey of such reductions.
See also \cite{MW95} for an account how reductions can be used as a
framework for classification, and for a survey of applications of
twistor theory.

 It has been shown \cite{MN89} that the SDYM equations with gauge group the
volume-preserving diffeomorphisms SDiff$({\cal M})$ of a four-manifold
${\cal M}$ and translational symmetry in all four variables reduces to
the self-dual (SD) Einstein vacuum equations on ${\cal M}$. 
This result extends the work of Ashtekar et al.\ \cite{AJS88}.  
It also implies, \cite{Wa90}, that solutions of the  SDYM equations
with two translational symmetries and
gauge group  SDiff$(\Sm)$ for some two-manifold $\Sm$ also
determine solutions of the SD Einstein vacuum equations.

 The aim of the present
paper is to show that the correspondence between the Lax formulations of
certain two-dimensional integrable systems and the SD Einstein
equations enables us to construct SD vacuum metrics explicitly
from solutions to various two-dimensional nonlinear integrable
equations.  We do this by considering $SL(2, \C)$ SDYM fields
invariant under the action of two translations of space-time.
These fields are can be represented as 
 solutions of various soliton equations in two
dimensions.  Self-dual vacuum metrics are recovered by 
representing the  Lie algebra of (real forms of) $SL(2, \C)$
 as Hamiltonian vector fields on a two-dimensional homogeneous
space for the gauge group.

Other approaches to self-dual gravity that reveal its connection with
two-dimensional integrable systems have  been
given by Ward \cite{Wa90} and Q-Han Park \cite{Ph90}.

In the next section we review briefly the classification of
two-dimensional integrable systems arising from the $SL(2, \C)$ SDYM
equations.  In section 3 we
discuss the connection between the SDYM equations  and self-dual
gravity. 
Section 4
is devoted to the construction of normalised null tetrads and hence
 metrics on $\R^2 \times \Sm$ from the SDYM Lax pairs for
the two-dimensional integrable systems.  In the last section we outline
the twistor interpretation of the construction.

\section{Self-dual Yang-Mills and 2D integrable systems} 
Consider a Yang Mills vector bundle over a four-dimensional manifold $\MMM$
(taken here
to be $\C^4$ in general, or $\R^4$ when reality conditions are
imposed) with connection one-form $A=A_{\mu}(x^{\nu}) dx^{\mu} \in
T^{\ast}\MMM\otimes LG$, where $LG$ is the Lie algebra of some gauge
group $G$.  The corresponding curvature $F=F_{\mu \nu}dx^{\mu}\wedge
dx^{\nu}$ is given by 
\be 
F_{\mu \nu}=[D_{\mu}, D_{\nu}]=
\p _{\nu}A_{\mu} -
\p_{\mu}A_{\nu}
+[A_{\mu}, A_{\nu}], 
\ee 
where $D_{\mu}=\p _\mu - A_{\mu}$ is the covariant derivative.  The SDYM
equations on a connection $A$ are the self-duality conditions on the
curvature under the Hodge star operation
\be
\label{fstar}
F=\ast F,\;\;\;\;{\mbox {or in index notation} }\;\;\; 
F_{\mu \nu}=1/2{\epsilon}_{\mu \nu \sigma \rho }F^{\sigma \rho}.
\ee
They  are conformally invariant and are  
also preserved by the gauge transformations
\be
\label{fullgauge}
A \rightarrow g^{-1} Ag - g^{-1}dg,
\;\;\;\;\;\; g\in Map(\MMM, G). 
\ee
Let us introduce double-null coordinates $\w , \wt , z, \zt$, 
in which the metric on $\MMM$ is
$ds^2=d\w d\wt - dzd\zt$.
In these coordinates the SDYM equations  may be rewritten as
\begin{eqnarray}
\label{sdym1}
F_{\w z} &=& 0\\
\label{sdym2}
F_{{\wt}{\zt}} &=& 0\\ 
\label{sdym3}
F_{{\w}{\wt}}-F_{z{\zt}} &=& 0,
\end{eqnarray}
which  are the
compatibility conditions $[L,M]=0 $ for the linear system of equations
$L{\Phi} =0$, $M{\Phi} =0$ where the `Lax pair', $L$ and $M$, are
\be
\label{paralaxa}
L=D_{\w}-{\lambda}D_{\zt},\;\;\;\;M=D_z-{\lambda}D_{\wt}
\ee
for $\lambda \in {\C P^1}$ and
$\Phi$ an $n$-component column vector. 

We shall consider the reality conditions for real ultra-hyperbolic
spaces, recovered by imposing $\w =x-y$, $z=t+v$, $\wt = x+y$, $\zt
=t-v$.  (Reality conditions for Euclidean space are recovered by
imposing $\wt=\bar{w}$ and $\zt=-\bar{z}$.) Solutions to
(\ref{sdym1}--\ref{sdym3}) can be real for this choice of signature.

We fix the gauge group to be $SL(2, \C)$ or one of its real subgroups.
Conformal reduction of the SDYM equations
involves the choice of the group $H$ of
conformal isometries of $\MMM$.
We shall restrict ourselves to the simplest case 
and suppose that a connection $A$ is invariant under the flows
of two independent translational Killing vectors $X$ and $Y$.
These reductions are classified partially by the signature of the
metric restricted to two-plane spanned by  the translations.

\begin{itemize}
\item[1)] {\bf Nondegenerate cases (${\bf H_1}$)}
\begin{enumerate} 
\item[a)]
$ X={\p}_{\w}-{\p}_{\wt},\; Y={\p}_z-{\p}_{\zt}$.
\begin{eqnarray}
\label{sgm}
A_{\w} = {\AU}  ,&A_{\wt} = {\AV}\nonumber\\
A_{z} = {\AZ}  ,&A_{\zt} = {\AW}  .
\end{eqnarray}
The SDYM equations are satisfied in ultra-hyperbolic signature if
${\ps}_{xx}+{\phi}_{tt}={\sin}{\phi}$; the elliptic sine-Gordon
equation.
\item[b)]
$G=SU(2),\;\; X={\p}_{\w},\; Y={\p}_{\wt}.$ 
\[
A_{\zt}=0,\;\;\;\;\;\;A_{\w}=cos{\phi}{\pau1}+sin{\phi}{\lam},
\]
\be
\label{sghm}
A_{\wt}={\pau1},\;\;\;\;\;\; A_{z}=1/2({\phi}_v-{\phi}_t){\lap}.  
\ee
The SDYM equations in ultra-hyperbolic signature yield
${\phi}_{tt}-{\phi}_{vv}=4{\sin}{\phi}$, the hyperbolic sine-Gordon
equation.
\end{enumerate}
For details of these reductions, see \cite{MW95} Chapter 6 or
\cite{AC92}.  Note also that if we reduce from Euclidean signature we
obtain Hitchin's Higgs bundle equations (which can also be represented
as harmonic maps from $\R^2$ to $SL(2,\C)/G$ where $G$ is $SU(2)$ or
$SU(1,1)$) \cite{MW95}.

\item[2)] {\bf Partially degenerate case (${\bf H_2}$)}\\ We consider
ultra-hyperbolic signature only with $X={\p}_{\w}-{\p}_{\wt}$ and 
$Y={\p}_{\zt}$.
\begin{enumerate}
\item[a)]
\be
\label{KdV}
A_{\w}={\Ak},\; A_{\wt}=0,\;2A_z={\Qk},\; A_{\zt}={\Bk},
\ee 
where $4w=q_{xxx}-4qq_{x}-2{q_x}^2+4q^2q_x$ and $b=q_x-q^2$.
The SDYM equations 
(with the definition $u=-q_x$) are equivalent to the  Korteweg de Vries 
equation $4u_z=u_{xxx}+12uu_x$. The reduced Lax pair (\ref{paralaxa})
yields the standard zero curvature representation of KdV \cite{NMPZ84}. 
\item[b)]
\be
\label{AC2}
A_{\w}={\A2},\;A_{\wt}=0,\;
A_z=i{\Q2},\; 2A_{\zt}=\pm i{\B2}
\ee
Here the upper (lower) sign corresponds to $ G=SU(2)$ 
(or $SU(1, 1)$).  
SDYM become  $i\ps _z=-\ps _{xx} \mp 2{\psm}^2 \ps $
which is the nonlinear Schr\"{o}dinger equation with an attractive (respectively repulsive)
 self interaction \cite{MS89}.
\end{enumerate}
\end{itemize}

\section{SDYM and self-dual gravity}
\newtheorem{thr}{Theorem}[section] Let $\cal M$ be a four-dimensional
complex manifold (for example the complexification of some real slice
${\mathcal{M}_{\R}}$) and let $g$ be a holomorphic metric on $\cal M$
(for example the complexification of a real metric on ${\mathcal{M}_{\R}}$).
The following theorem states that the self-duality equations on the
curvature can be expressed in terms of the consistency condition for a
Lax pair of vector fields.
\begin{thr}[Mason \& Newman 1989 \cite{MN89}].
\label{mn}Let $V_a=(W, \Wt, Z, \Zt )$ be four independent holomorphic vector
fields on a four-dimensional complex manifold ${\cal M}$ and let $\nu$
be a nonzero holomorphic 4-form. Put 
\be
\label{sde}
L=W-{\lambda}\Zt,\;\;\;\;\;\;\;M=Z-{\lambda}\Wt.
\ee
Suppose that for every $\lambda \in \C P^1$
\be
\label{podstawka}
[L, M]=0
\ee
\be
{\cal L}_L{\nu}=-{\cal L}_M{\nu}=0
\ee
Here ${\cal L}_V$ denotes a Lie derivative.
Then ${\sigma}_a = f^{-1}V_a$, where $f^2={\nu}(W, \Wt, Z, \Zt)$, is a normalised
null-tetrad for a half-flat metric (i.e.\ with vanishing Ricci tensor
and self-dual Weyl tensor). Every half-flat metric arises in this way.
\end{thr}

The covariant metric is conveniently expressed in terms of the dual frame
$e_{V_a}$
\be
\label{zero1}
g=f^2(e_W \odot e_{\Wt} - e_Z \odot e_{\Zt}), 
\ee
where
\begin{eqnarray}
\label{jednoformy}
e_W=f^{-2}\nu (..., \Wt, Z, \Zt)\; &,& e_{\Wt}=f^{-2}\nu (W, ..., Z,\Zt)\nonumber\\
e_Z=f^{-2}\nu (W, \Wt, ..., \Zt)\; &,& e_Z=f^{-2}\nu (W, \Wt, Z, ...).
\end{eqnarray}
The operators $L$ and $M$ determine a basis of ASD two-forms on $\cal M$
\be
\label{hyper}
\a =f^2e_W \wedge e_Z,\;\;\;  
\om =f^2(e_W\wedge e_{\Wt}-e_Z\wedge e_{\Zt}),\;\;\;
\at =f^2e_{\Wt}\wedge e_{\Zt}. 
\ee
We note that $-i(\a -\at ), i\om $ and $\a +\at$ are nondegenerate symplectic forms,
which (together with three compatible complex structures ) endow 
$\cal M$ with a complexified hyper-K{\"{a}}hler structure.

\section{Self-dual metrics on $\R^2 \times \Sm$}\label{sdmetrics2d}
We connect the self-duality equations on a Yang-Mills field and those
on a four-dimensional metric by considering gauge potentials that take
values in a Lie algebra of vector fields on some manifold. Theorem
(\ref{mn}) reveals one such connection: $ W, \Wt, Z$ and $\Zt$ are
generators of the group of volume-preserving (holomorphic)
diffeomorphisms of $( {\cal M},\, \nu )$. We make the identification:
$W=D_{\w},\; \Wt=D_{\wt},\; Z=D_z,\; \Zt=D_{\zt}$.  By comparing
(\ref{podstawka}) with (\ref{paralaxa}), we see that the half flat
equation is a reduction of the SDYM with this gauge group by
translations along the four coordinate vectors $\p _{\w},\; \p
_{\wt},\; \p _{z},\; \p _{\zt}$. 

In order to understand the relationship with two-dimensional
integrable systems, we look at this in a 
slightly different way.  Let $( \Sm ,\; \Om _{\Sm} )$ be a
two-dimensional symplectic manifold and let SDiff$( \Sm )$ be the group
of canonical transformations of $\Sm $.  Consider the SDYM equations 
with the gauge
group $G$, where $G$ is  the subgroup of SDiff$( \Sm )$. We can represent the
components of the connection form of $D$ by Hamiltonian vector fields and
hence by Hamiltonians on $\Sm $ depending also on the coordinates
on$\MMM$:
 \be W=\p
_{\w}-X_{H_{\w}},\; \Wt=\p _{\wt}-X_{H_{\wt}},\; Z=\p _z-X_{H_{z}}, \;
\Zt=\p _{\zt}-X_{H_{\zt}} \ee where $X_{H_{\mu}}$ 
denotes the
Hamiltonian vector field corresponding to $A_{\mu}$ with Hamiltonian
$H_\mu$.

Now we suppose that $D$ is invariant under two
translations. The reduced Lax pair will then descend to $\R^2 \times
\Sm$ and give rise to a half flat metric.  This requires that
the gauge group is a subgroup of the canonical transformations of $\Sm
$. Although it has been observed that SDiff$(\Sm )\approx SL(\infty
)$, it seems that $SL(n, \C)$ is a subgroup of such defined $SL(\infty
)$ only for $n=2$ \cite{M90}.  In this case we can take the linear
action of $SL(2, \R)$ on $\R^2$ or a M{\"{o}}bius action of $SU(2)$
and $SU(1, 1)$ on $\C P^1$ or $D$ (the Poincar\'e disc) respectively. We
shall restrict ourselves to real vector fields, which will imply that
our SD metrics will have ultra-hyperbolic signature (Euclidean
examples can also be obtained in a similar way).

To be more explicit we write down the Hamiltonian{\footnote{ We only
require the representation of $A_{\mu}$ by volume-preserving vector
fields on $\Sm$; Hamiltonians are defined up to the addition of a function
of the (residual) space variables, but different choices of such
functions do not change the metric.}}
corresponding to the matrix
\[ 
A_{\mu}=\AM \in LSL(2, \C).
\]
In the three cases we have
\be
\label{polelin}
\Sm =R^2,\;\;\os = dm \wedge dn,\;\; H_{\mu} = (\frac{bn^2}{2}+amn
-\frac{cm^2}{2}),
\ee
\be
\label{hampole}
\Sm =\C P^1,\;\; \os= \frac{id\xi \wedge d{\xib }}{{(1+\xi \xib )}^2},\;\; 
  H_{\mu}=-i\frac{\xi\ov{b}-\xib b +2a}{1+\xi \xib},
\ee
\be
\label{hamm}
\Sm =D,\;\;\os= \frac{id\xi \wedge d{\xib }}{{(1-\xi \xib )}^2},\;\;  
 H_{\mu}=-i\frac{\xi\ov{b}-\xib b -2a}{1-\xi \xib}.
\ee

The form of the null tetrad (\ref{jednoformy})
and the hyper-K\"{a}hler structure (\ref{hyper})  
obtained after the two-dimensional reductions of SDYM is as follows:

 {\bf(i) ${\bf H_1}\;\;(X=\p _{\w},Y=\p_{\wt}),\;\; 
\nu =dz \wedge d{\zt} \wedge {\Om}_{\Sm}$}
\be
\label{h11}
 f^2=\nu (W, \Wt , Z, \Zt)=\os (W, \Wt )=\{H_w, H_{\wt} \}= F_{w\wt}.
\ee 
In the last formula $F_{w\wt}$ is a function rather then a
matrix. This follows from the 
identification (via (\ref{polelin})-(\ref{hamm})) of $2\times 2$
matrices in the Lie algebra of $SL(2,\C)$ and Hamiltonians. 
Let $d_{\Sm}$ stand for the exterior derivative on $\Sm$.
\begin{eqnarray}
\label{h1tetrad}
e_W &=& f^{-2}(\os (\Wt, Z)dz +\os (\Wt, \Zt )d{\zt}
+\os (..., \Wt))\nonumber \\ 
&=& f^{-2}(\{ H_{\wt}, H_{z}\} dz+ \{ H_{\wt}, H_{\zt}\} d{\zt}
- d_{\Sigma}H_{\wt}) \nonumber\\
e_{\Wt} &=& f^{-2}(\{ H_{\w}, H_{z}\}dz -\{ H_{\w}, H_{\zt}\}d{\zt} 
+d_{\Sigma}H_{\w})\nonumber \\
e_Z &=& dz\\
e_{\Zt} &=& d{\zt}\nonumber
\end{eqnarray}
\begin{eqnarray*}
\a &=& -\{ H_{\wt}, H_{\zt}\} dz\wedge d\zt -\ds H_{\wt}\wedge dz \\ 
\om &=&  ( \{ H_{z}, H_{\zt}\} - \{H_w, H_{\wt} \}) dz\wedge d\zt 
+{\Om}_{\Sm}+\ds H_{z}\wedge d{z} +\ds H_{\zt}\wedge d{\zt}\\
\at &=& \{ H_{\w}, H_{z}\} dz\wedge d\zt 
+\ds H_{w}\wedge d{\zt} .
\end{eqnarray*}
The gauge freedom is used to set $A_{\zt}$ (and hence $H_{\zt}$) to $0$. 
\[
ds^2=\frac{1}{\{H_w, H_{\wt} \}}\Big(-(\{ H_{\wt}, H_{z}\}\{ H_{\w}, H_{z}\})
dz^2-{\{H_w, H_{\wt} \}}^2dzd\zt
\]     
\[
-(\p _{\xi}H_{\w}\p _{\xi}H_{\wt}){d\xi}^2
-(\p _{\xib}H_{\w}\p _{\xib}H_{\wt}){d\xib}^2
-((\p _{\xi}H_{\wt}\p _{\xib}H_{\w})+(\p _{\xib}H_{\wt}\p _{\xi}H_{\w}))d\xi 
d\xib
\]
\[
+(\p _{\xi}H_{\wt}\{ H_{\w}, H_{z}\}+\p _{\xi}H_{\w}\{ H_{\wt}, H_{z}\})dzd\xi
+(\p _{\xib}H_{\wt}\{ H_{\w}, H_{z}\}+\p _{\xib}H_{\w}\{ H_{\wt}, H_{z}\})
dzd\xib \Big).
\] 
{\bf(ii) ${\bf H_2},\;(X=\p_{\w}-\p _{\wt},Y=\p_{\zt}),\;\; 
\nu =dx \wedge dz \wedge {\Om}_{\Sm}, $}
\be
\label{h22}
f^2=\{ H_{\w}-H_{\wt}, H_{\zt} \}=F_{w\zt},
\ee
\begin{eqnarray}
\label{h2tetrad}
e_W  &=& f^{-2}(\{ H_{\zt}, H_{\wt}\} dx+ \{ H_{\zt}, H_{z}\} dz -
d_{\Sigma}H_{\zt})\nonumber\\
e_{\Wt} &=& f^{-2}(-\{ H_{\zt}, H_{\w}\} dx -\{ H_{\zt}, H_{z}\} dz +d_{\Sigma}H_{\zt})\nonumber \\
e_Z &=& dz\\  
e_{\Zt} &=&f^{-2}(\{ H_{\w}, H_{\wt}\} dx+ \{ H_{\w}-H_{\wt}, H_{z}\} dz -
d_{\Sigma}( H_{\w}-H_{\wt}))\nonumber
\end{eqnarray}
\begin{eqnarray*}  
\a &=& \{ H_{\zt}, H_{\wt}\} dx\wedge dz +\ds H_{\zt}\wedge dz\\ 
\om  &=& (\{ H_{\w}, H_{z}\} - \{ H_{\wt}, H_{\zt}\}) dx\wedge dz
+\ds H_{\zt}\wedge dx -\ds (H_{w}-H_{\wt})\wedge dz\\
\at  &=&  \{ H_{\w}, H_{z}\} dx\wedge dz+{\Om}_{\Sm}+\ds H_{w}\wedge dx 
-\ds H_{z}\wedge dz. 
\end{eqnarray*}
We can perform a further gauge transformation to set $H_{\wt}=0$ in which case
\[
ds^2=
-\left( \frac{ \{ H_{\zt}, H_{z}\}^2}{\{ H_{\w}, H_{\zt} \}}+\{ H_{\w}, H_{z}\}\right) dz^2
-\frac{(\p _{\xi}H_{\zt})^2}{\{ H_{\w}, H_{\zt} \}}{d\xi}^2
-\frac{(\p _{\xib}H_{\zt})^2}{\{ H_{\w}, H_{\zt} \}}{d\xib}^2
\]
\[
+\left( \p _{\xi} H_{\w} +2\frac{ \{ H_{\zt}, H_{z}\}}{\{ H_{\w}, H_{\zt} \}}
\p _{\xi} H_{\zt} \right) dzd{\xi}
+\left( \p _{\xib} H_{\w} +2\frac{ \{ H_{\zt}, H_{z}\}}{\{ H_{\w}, H_{\zt} \}}
\p _{\xib} H_{\zt}\right) dzd{\xib}
\]
\[
-\frac{2\p _{\xi}H_{\zt}\p _{\xib}H_{\zt}}{\{ H_{\w}, H_{\zt} \}}d\xi d\xib
+\{ H_{\zt}, H_{z}\}dxdz-\p _{\xi}H_{\zt}dxd\xi-\p _{\xib}H_{\zt}dxd\xib.
\]
Reductions by $X=\p _{\w},\; Y=\p _{z}$ are not considered because the
resulting metric turns out to be degenerate everywhere as a direct
consequence of the SDYM equations. Equation (\ref{sdym1}) becomes now $[X_{H_{\w}},
X_{H_z}]=0$ which, in the case of finite dimensional sub-algebras of
LSDiff$(\Sm )$, implies linear dependence of $X_{H_{\w}}$ and
$X_{H_z}$.

The construction naturally applies to the complex four-manifolds.
 We start from the SDYM equations
on $\C^4$ with gauge group $SL(2, \C)$. Then we perform one of the
possible reductions to $\C^2$. Let ${{\Sm}^2}_{\C}$ be a
two-dimensional complex manifold, for example $\C P^1 \times \C
P^{1\ast} $. $SL(2, \C)$ acts on one Riemann sphere by a
M{\"{o}}bius transformation, and the other by the inverse:
\[
(\xi ,\xit)\longrightarrow \left( \frac{A\xi +B}{C\xi +D},
\frac{D\xit - C}{-B\xit +A} \right) .
\]
Here $\xi$ and $\xit$ are independent complex coordinates on $\C P^1$
and $\C P^{1\ast}$. The action preserves the symplectic form $
\Om_{\Sm _{\C}}= (1+\xi \xit)^{-2}(d\xi \wedge d\xit)$ defined on the
complement of $1+\xi\xit=0$.  All results of this section may be extended
to the complex case by replacing $\xib$ by the independent coordinate
$\xit$.

\subsection{Solitonic metrics}
We can now establish the connection between the 
integrable systems reviewed in section 2 and self-dual vacuum
metrics. We do so be expressing the Hamiltonians above in terms of
solutions to various soliton equations.
\label{metrsol}
From a given solution of a two-dimensional nonlinear equation we can generate a 
null tetrad (\ref{jednoformy}).
\begin{itemize}
\item[1)]{\bf NlS}
\begin{eqnarray*}
\label{metnls}
W &=& \p _x+(\psb {\xi}^2+\ps)\p _{\xi} +(\ps {\xib}^2+\psb)\p _{\xib}\\
\Wt &=& \p _x\\
\Zt &=& -i\xi \p _{\xi}+i\xib \p _{\xib}\\
Z &=& \p _z +i(-\psb _x {\xi}^2 +2{\psm}^2\xi +\ps _x)\p _{\xi}  
-i(-\ps _x{\xib}^2 +2{\psm}^2\xib +\psb _x)\p _{\xib}\nonumber\\ 
f^2 &=& \frac {2Re(\xib \ps)}{1+{\xm}^2}   
\end{eqnarray*}
\item[2) ] ${\bf KdV}$
\begin{eqnarray*}
W &=& \p _x +(qm+n)\p _m +(bm-qn)\p _n\\
\Wt &=& \p _x\\
\Zt &=& m\p _n\\
Z &=& \p _z +(\frac{b_x}{2}m-q_xn)\p _m+(wm-\frac{b_x}{2}n)\p_n\\
f^2 &=&-m(q+mn)
\end{eqnarray*}
where $b=q_x-q^2$ and $ 4w=q_{xxx}-4qq_{xx}-2{q_x}^2+4q^2q_x$.

\item[3)]${\bf SG}$; elliptic case.
\begin{eqnarray*}
W &=& {\p}_x+\frac{1}{4}(\ps _tm-2\cos (\ps /2)n){\p _m} 
+ \frac{1}{4}(-\ps _tn-2\cos (\ps /2)m){\p _n}\\ 
\Wt &=& {\p}_x+\frac{1}{4}(\ps _tm+2\cos (\ps /2)n){\p _m} 
+ \frac{1}{4}(-\ps _tn+2\cos (\ps /2)m){\p _n}\\ 
\Zt &=& {{\p}_t}+\frac{1}{4}(-\ps _xm-2\sin{(\ps /2)}n){\p _m} 
+ \frac{1}{4}(\ps _xn-2\sin{(\ps /2)}m){\p _n}\\ 
Z &=& {{\p}_t}+\frac{1}{4}(-\ps _xm+2\sin{(\ps /2)}n){\p _m} 
+ \frac{1}{4}(\ps _xn+2\sin{(\ps /2)}m){\p _n}\\
f^2 &=& (\sin {\ps})mn
\end{eqnarray*}
\item[4)]${\bf  SG}$; hyperbolic case
\begin{eqnarray*}
W &=& (-i{\xi}^2e^{-i\ps }+ie^{i\ps })\p _{\xi}+
(i{\xib}^2e^{i\ps}-ie^{-i\ps })\p _{\xib}\\
\Wt &=& (-i{\xi}^2+i)\p _{\xi}+(i{\xib}^2-i)\p _{\xib}\\
\Zt &=& \p _{\zt}\\
Z &=& \p _z -i(\p _{\zt}\ps )\xi \p _{\xi} 
	    +i(\p _{\zt}\ps)\xib \p _{\xib}\\   
f^2 &=& \frac{4\sin {\ps}({\xm}^2-1)}{{\xm}^2+1}.
\end{eqnarray*}
\end{itemize}
Put $d_A\xi =d\xi +i\xi {\p _{\zt}}\ps\; dz$. Then we have
\[   
ds^2= \frac{1}{1+\xi \xib}\Big( [ (1-{\xib}^2)^2\cot{\ps} + i(1-{\xib}^4)]
d_A\xi \otimes d_A\xi +2\sin{\ps}\;dz\otimes d\zt 
\]
\be
\label{sinemetric}
+( \cot{\ps} (1-{\xib}^2)(1-{\xi}^2)+i[(1+{\xib}^2)(1-{\xi}^2)
\ee
\[
 -(1-{\xib}^2)(1+{\xi}^2)] )  {d_A}\xi \otimes \overline{d_A}\xib
+[ (1-{\xi}^2)^2\cot{\ps} - i(1-{\xi}^4)]
 \overline{d_A}\xib \otimes \overline{d_A}\xib \Big).
\]
If one takes a solution describing the interaction of a half
kink and a half anti-kink (two topological solitons travelling in $z-\zt$ 
direction and increasing from $0$ to $\pi$ as $z+\zt$ goes from $-\infty$ to
$\infty$) then the singularity in $\sin{\ps}=0$ may be absorbed by a conformal
transformation of $z+\zt$ \cite{D96}.

From the Yang-Mills point of view,  the solutions that we have
obtained are  metrics on the total space of 
${\cal E}$, the  $\Sm$-bundle  associated to the Yang-Mills bundle.
Therefore it is of interest to consider the effect of gauge transformations.
First notice that diffeomorphisms of $\R^2 \times \Sm$ given by
\be
\label{sympdiff}
x^a\longrightarrow x^a + \epsilon X_F(x^a) \ee yield
$H_{\mu}\longrightarrow H_{\mu}+\epsilon (\{ H_{\mu}, F \} +\p
_{\mu}F)$ which is an infinitesimal form of the full gauge
transformation (\ref{fullgauge}). Here $\mu$ is an index on $\MMM$,
whereas $a$ is an index on ${\cal M}=\R^2 \times \Sm$. The vector field
$X_F$ is Hamiltonian with respect to $\Om_{\Sm}$, with Hamiltonian  
$F=F(x^a)$.

If (\ref{sympdiff}) preserves the
K{\"{a}}hler structure of $\Sm $ then $H_ {\mu}$ transforms under
(a real form of) $SL(2, \C)$ and therefore our construction remains
`invariant'.

\section{Final remarks}
\subsection{The relationship between the twistor correspondences}
To finish, we explain how our construction ties in with the twistor
correspondences for the self-duality equations.  We consider only the
complex case of the SDYM equations  with two commuting symmetries $X, Y$. The $SL(2,
\C)$ SDYM connection defines, by the Ward construction \cite{Wa77}, a
holomorphic vector bundle over the (non-deformed) twistor space,
${E_{W}\rightarrow\cal P}$.  It is convenient\footnote{The diagram
(\ref{diagram}) describes also the general case of
$G=SDiff({\Sm^2}_{\C})$.  For this we work with ${\cal E}_{W}^5$
rather than the principal Ward bundle, since the latter has
infinite-dimensional fibres.  The notation is such that the upper
index of a space stands for the complex dimension of that space.}  to
use the bundle ${\cal E}_{W}^5$ - associated to $E_{W}$ by the
representation of $SL(2, \C)$ as holomorphic canonical transformations
of the complex symplectic manifold ${\Sm^2}_{\C}$. 

On the other hand, the SD vacuum metric corresponds to a deformed
twistor space ${\cal P}_{{\cal M}}$, \cite{Pe76}. In this paper we
have explained how the quotient of ${\cal E}$ by lifts of $X, Y$ is,
by theorem (\ref{mn}), equipped with a half-flat metric .  To give a
more complete picture we can obtain the deformed twistor space
directly from ${\cal E}_{W}^5$ and show that this is the twistor space
of ${\cal M}$.  Consider the following chain of correspondences:
$$    
\begin{array}{cccccccccc}
 &{\cal E}_{W}^5 & &{\cal F}& &{\cal E}& &{\cal F}_{\cal M}& \\
\swarrow& &\searrow\;\swarrow& &\searrow\;\swarrow& &\searrow\;\swarrow& &\searrow\\
{\cal P}_{{\cal M}}& &{\cal P}& &\C^4& &{\cal M}& &{\cal P}_{{\cal M}}
\end{array}
$$
Here ${\cal F}$ and ${\cal F}_{\cal M}$ are the standard projective
spin bundles fibred over $C^4$ and $\cal M$ respectively.  The space
${{\cal F}_{\cal E}^7}$, the pullback of the spin bundle ${\mathcal
F}$ to the total space of the bundle ${\cal E}$, fibres over all the
spaces in the above diagram.  Taking the quotient by lifts of $X, Y$ we
project ${{\cal F}_{\cal E}^7}$ to ${\cal F}_{\cal M}$. Taking the quotient by the twistor
distribution, ${{\cal F}_{\cal E}^7}$ also projects to the Ward bundle
${\cal E}_{W}^5$. By definition it projects to ${\cal E}$ and it could
equivalently have been defined as the pullback of ${\cal E}$ to ${\cal
F}$.  The compatibility of these projections is a consequence of the
commutativity of the diagram
\begin{eqnarray}
\label{diagram}
\C P^1\times \C^4\times {\Sm}_{\C}^2={\cal F}_{\cal E}^7 
&\stackrel{(X, Y)}\longrightarrow &{\cal F}_{\cal M}^5\nonumber\\
\Big\downarrow & &\Big\downarrow\\
{\cal E}_{W}^5 &\stackrel{(X, Y)}\longrightarrow &{\cal P}_{{\cal M}}.\nonumber
\end{eqnarray}
which follows  from the integrability the the distribution spanned by 
(lifts of) $X, Y, L, M$, and from the fact that
$(X, Y)$ commute with $(L, M)$.

\subsection{Global issues}
In order to obtain a compact space one might attempt the following:
\begin{itemize}
\item choose the gauge group to be $SU(2)$ so that the fibre space is
compact, and
\item  Compactify $\R^2$ after the reduction.
\end{itemize}
We restrict the rate of decay of $A_{\mu}$ by the requirement
that $A_{\mu}$ should be smoothly extendible to $S^2$ in the split
signature case. Other possibilities are to 
restrict to the class of rapidly decreasing soliton
solution of corresponding integrable equation.  
If we have reduced from a Euclidean signature solution to the SDYM
equations, then it is more natural to compactify $\R^2$ in such a way
as to obtain a Riemann surface of genus greater than one as it is only
for such a compactifications that one can have existence of nontrivial
solutions, \cite{Hi87}.

However, we still have singularities in the metrics corresponding to
(\ref{h1tetrad}) and (\ref{h2tetrad}), even if we can eliminate those
from the Yang-Mills connection.  We are left with
singularities associated with sets on which the tetrad becomes
linearly dependent.  This reduces to the proportionality (or
vanishing) of the Higgs fields on $\Sm$, which generically occurs on a
real co-dimension one subset of each fibre (and hence co-dimension one
in the total space). In the above formulae this set is given by the
vanishing of $f$.  The Weyl curvature $C_{abcd}$ blows up as $f$ goes
to zero.
Calculation of curvature invariants show that these
lead to genuine singularities that cannot be eliminated by a change of
frame or coordinates. For example
\[
C_{abcd}C^{abcd}=\sum_{i=-3}^{3}C_if^{2i},
\]
where $C_i=C_i(x^a)$ are generally non-vanishing regular functions on
$\cal M$, which explicitly depend on Yang-Mills curvature $F_{\mu \nu}$ and
(derivatives of) Hamiltonians (\ref{polelin}-\ref{hamm}). Those 
singularities appear (for purely topological reasons) because
each vector in the tetrad $(W, \Wt, Z, \Zt )$ has 
at least one zero, when restricted to $\Sm=S^2$.

One can also obtain Euclidean metrics as above by using reductions of
the SDYM equations from Euclidean space, but we will still be unable to avoid these
same co-dimension one singularities.

\subsection{Other reductions}
We have focused in this article on the familiar $1+1$ soliton
equations.  However, it is clear from the discussion of section \S
3--\ref{sdmetrics2d} that the construction will extend to any symmetry
reduction of the SDYM equations to systems in two dimensions with gauge group contained in
$SL(2,\C)$, in particular when the symmetry imposed consists of two
translations as for the Euclidean signature examples mention
previously.  However, one can also use the same device to embed
examples using any other two-dimensional symmetry subgroup of the
conformal group.  In particular, with cylindrical symmetry, one
obtains the Ernst equations (the two symmetry reduction of the full,
non self-dual four-dimensional Einstein vacuum equations) and this can
similarly be embedded into the self-dual (but not vacuum) equations.

\section {Acknowledgements}
We thank  Maciej Przanowski for helpful comments on the manuscript. 

\end{document}